\documentstyle[romp32]{article}
\title{Region Operators of Wigner Function: Transformations, Realizations and Bounds}
\author{ Demosthenes Ellinas$^*$ and Ioannis Tsohantjis$^{**}$\\ Department of Sciences, Division of Mathematics$^*$ and Physics$^{**}$ \\
Technical University of Crete GR-731 00 Chania Crete Greece \\
$^*$ellinas@science.tuc.gr\qquad $^{**}$ioannis2@otenet.gr\\[2ex]}

\begin{document}

 \maketitle{\it Dedicated to Professors Angas Hurst and Tony
Bracken

on the occasion of their birthdays

(Reports of Mathematical Physics, to appear Feb. 2006)} \vskip
0.5cm

\begin{abstract}
     An integral of the Wigner function  of
a wavefunction $|\psi >$, over some region $S$ in classical phase
space is identified as a (quasi) probability measure (QPM) of $S$,
and it can be expressed by the $|\psi >$ average of an operator
referred to as the $\textit{ region operator }$(RO).
Transformation theory is developed which provides the RO for
various phase space regions such as point, line, segment, disk and
rectangle, and where all those ROs are shown to be interconnected
by completely positive trace increasing maps. The latter are
realized by means of unitary operators in Fock space extended by
$2D$ vector spaces, physically identified with finite dimensional
systems. Bounds on QPMs for regions obtained by tiling with discs
and rectangles are obtained by means of majorization theory.
\end{abstract}


\section{Introduction}
The Wigner quasidistribution  function on phase space $W(p,q)$
\cite{schleich}, provides an important tool in the description of
quantum systems from both theoretical and experimental point of
view. Of particular interest is for example, its application to
the field of quantum tomography \cite{leo},\cite{sch}. Although
the Wigner function shares several properties with classical
distribution functions on phase space it is not positive-definite.
As a consequence, the value of its integral, (to be called
alternatively quasiprobability integral(qpi), mass (qpm), or
volume (qpv)), over subregions of phase space, lies in general
outside the interval $[0,1]$, and may admit large positive or
negative values. The upper and lower bounds of such integrals
define tests e.g on the accuracy of experimental determinations of
the Wigner function in the sense that they give the degree to
which the value of the integral lies outside $[0,1]$, and thus
their determination is of particular importance. In \cite{bdw},
\cite{bew1}, \cite{bew2} and \cite{james} it has been shown that
such bounds, for given regions or contours of phase space for 1D
quantum systems, are determined by the maximal and minimal
eigenvalues of corresponding Hermitian operators $K_S$, called
region operators (RO), whose spectral analysis has been shown to
have a rich algebraic structure. Consequently the study of the
properties of such operators defined by
\begin{eqnarray}
K_S=\int_{\Gamma}\chi_{S}(\alpha)W(\alpha)da\;,
\end{eqnarray}
where $S$ is a subregion in phase space $\Gamma$,
$\alpha\in\Gamma$, $\chi_{S}(\alpha)$ is the characteristic
function corresponding to $S$ and $W(\alpha)$ are the Wigner
operators, is of great importance.

This article presents a detail analysis of transformation theory
for RO, and in particularly investigates the cases of the point,
line, rectangle and disk operators. The development of the theory
is mainly based on the notion of completely positive trace
increasing map (CPTI) (c.f \cite{kr}), which provides the
interconnection between various region operators. The application
of such maps is also shown to describe tiling processes in phase
space. Moreover the eigenvalues of the obtained ROs from step to
step during the tiling are shown to respect majorization
relations\cite{mo}. This allows the determination of upper and
lower bounds on qpi over regions in phase space obtained by a
tiling process.

An outline of the work is as follows: section two focuses on line
and rectangle ROs. The spectral analysis of straight line segment
ROs is presented and further it is shown that straight line ROs
are projection operators. On the other hand the normally ordered
form of rectangle ROs is obtained, and moreover it is shown that
all rectangle ROs obtained from an initial one by rigidly shifting
its position in the phase plane, are isospectral and thus have the
same quasiprobability volume.

In section three an operational construction of quasiprobability
measures of canonical polygon ROs is developed by introducing
appropriate completely positive trace increasing (CPTI) maps. An
explicit example is then given in which a general canonical
hexagon RO is constructed from the corresponding RO of an
isosceles triangle, by means of CPTI maps.

Finally in section four, appropriate CPTI maps are used to realize
a tiling process in phase space acting on rectangle and disk ROs
used as building blocks. The step matrix which connects the
eigenvalue-vectors of ROs at each step of the tiling process is
shown to be simply determined by doubly stochastic matrices. This
further allows to determine simple ordering relations between the
eigenvalues of successive ROs by means of the majorization theory.
This fact provides intervals of extremal values for the
eigenvalues i.e the qpm's, of the sequence of region operators, in
a tiling process. Section five briefly summarizes results and
outlines some prospects of the theory of region operators.

\section{ Point, Line and Rectangle Region Operators}

\begin{theorem}{Proposition}\label{TH1}
Region operators having support on a straight line segment of
length $L$ in angle $\theta$ with the $q-$axis read,
$K_{L}^{\theta}=\frac
{\sin[(Q\cos\theta+P\sin\theta)L]}{Q\cos\theta+P\sin\theta}\Pi,$
and satisfy the eigenvalue problem
$K_{L}^{\theta}|\Psi_{\pm}^{\theta}\rangle=\pm
\frac{2\sin(q_{\theta}L)}{q_{\theta}}|\Psi_{\pm}^{\theta}\rangle,$
with eigenvectors
$|\Psi_{\pm}^{\theta}\rangle=|q_{\theta}\rangle\pm|-q_{\theta
}\rangle,$ where $\{|q_{\theta}\rangle,$
$q_{\theta}\in\mathbf{R\}},$ is a complete set of vectors
determined in the number state basis by means of the
Hermite polynomials as $\langle q_{\theta}|n\rangle=\frac{1}{\pi^{1/4}}%
\frac{1}{2^{n/2}(n!)^{1/2}}\exp(-q_{\theta}^{2}/2)H_{n}(q_{\theta}%
)\exp(-in\theta).$
\end{theorem}
\textit{Proof:} Let $K_{0}$ be the region operator with support on
the point at the origin ("point operator", see fig. 1a),
identified with
the parity operator $\Pi$, since \\
$K_{0}=\int_{R^{2}}\delta(q)\delta(p)D(q,p)\Pi
D(q,p)^{\dagger}dqdp=\Pi$. Let us introduce a symmetric broadening
of the point at the origin along the $q-$axis of extension $L$, by
means of the positive map $K_{L}^{P}
=\varepsilon_{L}^{P}(K_{0})=\varepsilon_{L}^{P}(\Pi),$ constructed
from the displacement operator $D(q,p)=\exp i(pQ-qP)=\exp(\alpha
a^{\dagger}-a^{\ast}a),$ with $\alpha =\frac{1}{\sqrt{2}}(q+ip)$ and
$a=\frac{1}{\sqrt{2}}(Q+iP),$ $a^{\dagger
}=\frac{1}{\sqrt{2}}(Q-iP),$ to get
\begin{eqnarray}
K_{L}^{P} =\int_{\frac{-L}{2}}^{\frac{L}{2}}D(q,0)\Pi
D(q,0)^{\dagger}dq
=\int_{\frac{-L}{2}}^{\frac{L}{2}}D(2q,0)dq\Pi=\frac{\sin(PL)}{P}\Pi.
\end{eqnarray}
In this way we obtain the region operator $K_{L}^{P},$ with
support on the straight line segment of length $L$, symmetric with
respect to the origin and lying along the position axis, (see
figure 1b). Similarly for the segment along the momentum axis \
the \ operator is $\
K_{L}^{Q}=\varepsilon_{L}^{Q}(K_{0})=\frac{\sin(QL)}{Q}\Pi$. Since
$\Pi|q\rangle=|-q\rangle,$ we get $K_{L}^{Q}|\Psi_{\pm}^{Q}\rangle
=\pm\frac{2\sin(qL)}{q}|\Psi_{\pm}^{Q}\rangle,$ where $|\Psi_{\pm}^{Q}%
\rangle=|q\rangle\pm|-q\rangle,$ $q\in\mathbf{R}.$ In the same
manner for the region operator with support a on length $L$ line
segment centered at the origin and extended along the $p-$ axis,
we get $\ K_{L}^{P}|\Psi_{\pm}^{P}\rangle
=\pm\frac{2\sin(pL)}{p}|\Psi_{\pm}^{P}\rangle,$ where $|\Psi_{\pm}^{P}%
\rangle=|p\rangle\pm|-p\rangle,p\in\mathbf{R}.$ The
$|\Psi_{\pm}^{Q}\rangle$ generalized eigenvectors, have respectively
the following position and momentum representation: $
\psi_{\pm}^{Q}(q^{^{\prime}})=\langle q^{^{\prime}}|\Psi_{\pm}^{Q}%
\rangle=\delta(q^{^{\prime}}+q)\pm\delta(q^{^{\prime}}-q)$,\qquad
and  $\psi_{+}^{Q}(p^{^{\prime}})=\langle
p^{^{\prime}}|\Psi_{+}^{Q}\rangle=\frac
{1}{\sqrt{2\pi}}\cos(qp^{^{\prime}}),\qquad\psi_{-}^{Q}(p^{^{\prime}%
})=\langle p^{^{\prime}}|\Psi_{-}^{Q}\rangle=\frac{i}{\sqrt{2\pi}}%
\sin(qp^{^{\prime}})$.

Similarly for the position and momentum representations of the
generalized eigenvectors $|\Psi_{\pm}^{P}\rangle$ we
correspondingly obtain the functions \
$\psi_{+}^{P}(q^{^{\prime}})=\langle
q^{^{\prime}}|\Psi_{+}^{p}\rangle=\frac
{1}{\sqrt{2\pi}}\cos(pq^{^{\prime}})\ , \psi_{-}^{P}(p^{^{\prime}
})=\langle q^{^{\prime}}|\Psi_{+}^{P}\rangle=\frac{i}{\sqrt{2\pi}}
\sin(pq^{^{\prime}})$,\qquad and \
$\psi_{\pm}^{P}(p^{^{\prime}})=\langle
p^{^{\prime}}|\Psi_{\pm}^{P}
\rangle=\delta(p^{^{\prime}}+p)\pm\delta(p^{^{\prime}}-p)$.

If we now rotate by angle $\theta$ the line segment support of
region operator $K_{L}^{Q}$, by the rotation operator $e^{i\theta
N},$ where $N$ \ the number operator, we end up with the operator
\begin{equation}
K_{L}^{\theta}=e^{i\theta N}K_{L}^{Q}e^{-i\theta N}=\frac{\sin[(Q\cos
\theta+P\sin\theta)L]}{Q\cos\theta+P\sin\theta}\Pi.
\end{equation}
As special cases we obtain the region operators for the line segment along the
$p-$ axis $K_{L}^{\theta=0}=K_{L}^{Q},$ and along the $q-$ axis $K_{L}%
^{\theta=\pi/2}=K_{L}^{P}$\rule{5pt}{5pt}

\textit{Remarks}: 1) As the trace of a region operator equals the
area of the support of the operator itself, the two positive maps
$\varepsilon_{L}^{Q,P}:$ $K_{0}\rightarrow K_{L}^{Q,P},$
introduced previously to get the straight line segment operator
from a point operator are trace increasing maps i.e
$Tr(K_{L}^{Q})=Tr(\varepsilon_{L}^{Q}(K_{0}))=LTr(K_{0}).$

2) The region operator $K_{L}^{\theta}=\frac{\sin(Q_{\theta}L)}{Q_{\theta}}%
\Pi,$ is written in terms of the rotated operator $Q_{\theta}=Q\cos
\theta+P\sin\theta.$ Two copies of the latter with their angles of rotation
differing by $\pi/2,$
\begin{equation}
Q_{\theta}=Q\cos\theta+P\sin\theta,\qquad
Q_{\theta+\pi/2}=Q\sin\theta -P\cos\theta,
\end{equation}
are canonical i.e $[Q_{\theta},Q_{\theta+\pi/2}]=i\mathbf{1.}$

3) Some additional properties of the family of vectors
$\{|q_{\theta}\rangle,$ $q_{\theta}\in\mathbf{R\},}$ are: it is a
complete set i.e $\int_{-\infty
}^{\infty}dq_{\theta}|q_{\theta}\rangle\langle
q_{\theta}|=\mathbf{1,}$ its generalized vectors are  orthogonal
i.e $\langle q_{\theta}|q_{\theta
}^{^{\prime}}\rangle=\delta(q_{\theta}-q_{\theta}^{^{\prime}}),$
and can be
constructed from the zero number state as%
\begin{equation}
|q_{\theta}\rangle=\frac{1}{\pi^{1/4}}\exp\left[  -\frac{1}{2}q_{\theta}%
^{2}+\sqrt{2}\exp(i\theta)q_{\theta}a^{\dagger}-\frac{1}{2}\exp(2i\theta
)a^{\dagger2}\right]  |0\rangle,
\end{equation}
and so the parity operator flips the sign of its argument i.e
$\Pi|q_{\theta}\rangle=|-q_{\theta}\rangle.$ These same states are
also specified from their overlap with a coherent state i.e
\begin{equation}
\langle q_{\theta}|a\rangle=\frac{1}{\pi^{1/4}}\exp\left[  i\langle
Q_{\theta+\pi/2}\rangle q_{\theta}\right]  \exp[-\frac{1}{2}(q_{\theta
}-\langle Q_{\theta}\rangle)^{2}]\exp[-\frac{i}{2}\langle Q_{\theta}%
\rangle\langle Q_{\theta+\pi/2}\rangle],
\end{equation}
where $\langle
Q_{\theta}\rangle=\langle\alpha|Q_{\theta}|\alpha\rangle=\frac
{1}{\sqrt{2}}[\alpha\exp(-i\theta)+\alpha^{\ast}\exp(i\theta)]$.\\
4) Region operators with support on straight line segments \ with
proportional
lengths, in any direction, are commuting i.e $K_{\lambda L}^{\theta}%
K_{L}^{\theta}=K_{L}^{\theta}K_{\lambda L}^{\theta},$ for $\lambda\geq1,$ and
$0\leq\theta<2\pi.$

\begin{theorem}{Proposition} \label{TH2} Region operators having support
on a straight line are projection operators.
\end{theorem}

\textit{Proof:} Let the region operator $K_{0}=\Pi$. Its smearing
along the whole momentum $p-$axis results into the following
region operator%
\begin{eqnarray}
K_{Q=0}  =\int_{-\infty}^{\infty}dpD(0,p)\Pi
D(0,p)^{\dagger}=\frac{1}{\sqrt{\pi}}:e^{-Q^{2}}:=|q=0\rangle\langle
q=0|.
\end{eqnarray}
If we next transform this operator to have support along an axis
parallel to $p-$axis and crossing the position axis in the point
$q\in\mathbf{R}$ then we will obtain
\begin{eqnarray}
K_{Q=q}=D(q,0)K_{Q=0}D(q,0)^{\dagger}=\frac{1}{\sqrt{\pi}}:e^{-(q-Q)^{2}}:=|q\rangle\langle
q|,
\end{eqnarray}
namely the projection operator in position states. Similarly by
smearing the point operator along the position $q-$ axis we obtain
\begin{eqnarray}
K_{P=0} =\int_{-\infty}^{\infty}dq D(q,0)\Pi
D(q,0)^{\dagger}=\frac{1}{\sqrt{\pi}}:e^{-P^{2}}:=|p=0\rangle\langle
p=0|.
\end{eqnarray}
\bigskip This operator meets the $p-$axis at its zero point, so by displacing
it arbitrarily by $\ p\in\mathbf{R}$ we will obtain%
\begin{eqnarray}
K_{P=p}=D(0,p)K_{P=0}D(0,p)^{\dagger}=\frac{1}{\sqrt{\pi}}:e^{-(p-P)^{2}}:=|p\rangle\langle
p|,
\end{eqnarray}
namely the projection operator in momentum states. In the last
four equations above the expression of respective region operators
has been given also in normal ordered form i.e when the canonical
creation operator is placed in the left of the annihilation
operator.

 So far operators $K_{Q=q}=|q\rangle\langle q|,$ and
$K_{P=p}=|p\rangle\langle p|,$ which are projective and have
support on any straight line parallel to the $p-$axis crossing the
$q-$axis at the point $q,$ and respectively on any straight line
parallel to the $q-$axis crossing the $p-$axis at the point $p$
have been constructed. \ In order to rotate these support axes by
any desired angle $\theta$ with respect to e.g the $q-$ axis, we
should perform an additional unitary rotation with $e^{i\theta
N},$ where $N$ is the number operator, the generator of rotations
around the origin. Namely, we should consider the transformation
$K_{Q=q}$ $(K_{P=p})\rightarrow e^{i\theta N}K_{Q=q}e^{-i\theta
N}$ $(e^{i\theta N}K_{P=p}e^{-i\theta N}).$ The latter will not
change the
projective character of $K_{Q=q}=|q\rangle\langle q|,$ and $K_{P=p}%
=|p\rangle\langle p|,$ operators\rule{5pt}{5pt} (see also
\cite{woot})

\textit{Remarks}: 1) The region operator having support on the
whole plane is the unit operator. In order to obtain the region
operator on the plane we should "add" the region operators on all
lines parallel to any given line. Let e.g
$K_{Q=q}=|q\rangle\langle q|,$ and $K_{P=p}=|p\rangle\langle p|,$
the two projective operators that have support on some straight
line parallel to $p-$axis crossing the $q-$axis at the point $q,$
and respectively on some straight line parallel to the $q-$axis
crossing the $p-$axis at the point $p.$ Their integrals
$\int_{R}K_{Q=q}dq=\int_{R}|q\rangle\langle q|dq=\mathbf{1},$ and
$\int_{R}$ $K_{P=p}dp=\int_{R}|p\rangle\langle p|dp=\mathbf{1,}$
give the region operator supported on the plane, which actually is
the unit operator due to the completeness of position or
respectively momentum states. The same result would have been
obtained if we would had started with any line in angle $\theta$
with the $p,q$ axes.

2) Various region operators can easily be introduced by combining
the operators constructed above e.g a region operator with support
on a bundle of $n$
parallel lines crossing the $q-$axis at the points $P_{n}=\{q_{1}%
,q_{2,}...,q_{n}\}$ is defined as $K_{P_{n}}=|q_{1}\rangle\langle
q_{1}|+|q_{2}\rangle\langle q_{2}|+...+|q_{n}\rangle\langle
q_{n}|.$ The latter can be obtained acting on
$|q_{1}\rangle\langle q_{1}|,$ the operator with support on the
first line of the bundle, with the trace increasing map
$\varepsilon_{n},$ i.e $\
K_{P_{n}}=\varepsilon_{n}(|q_{1}\rangle\langle
q_{1}|)=\sum_{i=1}^{n}V_{i}|q_{1}\rangle\langle
q_{1}|V_{i}^{\dagger},$ which has an operator sum representation
in terms of the unitary operators $V_{i} =\exp(i(q_{i}-q_{1})P),$
$i=1,2,...,n.$

Let us now show a more special result related to motions that can
be done on rectangle region operators that leave their spectrum,
namely their associated quasiprobability volumes, invariant.


\begin{theorem}{Proposition} \label{TH3}
If we denote by $\ K_{S}\equiv K(x_{0},k_{0};A,B)$ the region
operator with support on a rectangle domain $S$ in phase space
with lower left corner $\ $at the point $(x_{0},k_{0})$ and sides
of lengths $A$ and $B$ along the axes of position and momentum
variables respectively, then i) the normal order form of this
operator is given by equation (\ref{no}) ii) all \ operators
$K(x_{0},k_{0};A,B),$ $K(x_{0}+s,k_{0}+t;A,B)$ for $s,t$ $\in R$
are isospectral, namely they are region operators obtained from
the initial rectangle operator $K(x_{0},k_{0};A,B)$ by rigidly
shifting its position by $s,t$, and have all eigenvalues equal,
which then implies that they have the same quasiprobability
volumes.
\end{theorem}

\textit{Proof }: i) Let us recall that in general an operator
$K_{S}$ associated to a phase space region $S$ is given in terms
of the Wigner operator
\begin{equation}
W(\alpha)=\frac{1}{2\pi}D(\alpha)(-1)^{N}D(\alpha)^{\dagger}%
=\frac{1}{2\pi}e^{i\pi(a^{\dagger}-\alpha^{\ast})(a-\alpha)}=\frac{1}{2\pi
}:e^{-2(a^{\dagger}-\alpha^{\ast})(a-\alpha)}:
\end{equation}
The last equation is in normally ordered form and it is easily
obtained by means of the formula
$e^{\lambda(a^{\dagger}-\alpha^{\ast})(a-\alpha
)}=:e^{(e^{\lambda}-1)(a^{\dagger}-\alpha^{\ast})(a-\alpha)}:$
valid for complex parameter $\lambda.$ Then by means of the
(over)complete basis of coherent state vectors
$|z=q+ip\rangle\equiv|qp\rangle,$ $z\in C,$ the region operator
becomes
\begin{eqnarray}
K_{S}=\frac{1}{2\pi}\int_{z\in C}\int_{\alpha\in\sup p(\chi_{S}%
)}:e^{-2(a^{\dagger}-\alpha^{\ast})(a-\alpha)}d^{2}\alpha : \nonumber\\
=\frac{1}{(2\pi)^{2}}:\int_{z\in C}\int_{\alpha\in\sup p(\chi_{S}%
)}e^{-2(z^{\ast}-\alpha^{\ast})(z-\alpha)}|z\rangle\langle z|d^{2}%
\alpha\;d^{2}z:.
\end{eqnarray}

At this point we employ the technique of integration within an
ordered product (IWOP)\cite{iwop}, which applies in our case since
the above operator integrand \ is in normally ordered form. In
effect splitting the integration and applying IWOP yields, after
using the integral representation of error function
$erf(x)=\frac{2}{\sqrt{\pi}}\int_{0}^{x}e^{-t^{2}}dt$ , the normal ordered expression of the region operator%
\begin{eqnarray}
K_{S} =\frac{1}{(2\pi)^{2}}:\int_{(q,p)\in R^{2}}(\int_{x_{0}}^{x_{0}%
+A}e^{-(q-x)^{2}}dx)(\int_{k_{0}}^{k_{0}+B}e^{-(p-k)^{2}}dk)\;|qp\rangle
\langle qp|dqdpdxdk:\\
=\frac{1}{16\pi}:\left[erf(-\frac{1}{\sqrt{2}}(a^{\dagger
}+a)+x_{0}+A)-erf(-\frac{1}{\sqrt{2}}(a^{\dagger}%
+a)+x_{0})\right]  \times\nonumber\\
\left[  erf(-\frac{i}{\sqrt{2}}(a^{\dagger}-a)+k_{0}%
+B)-erf(-\frac{i}{\sqrt{2}}(a^{\dagger}-a)+k_{0})\right]
:\label{no}.
\end{eqnarray}

ii) If we rewrite the rectangle operator $K_{S}\equiv
K(x_{0},k_{0};A,B)$ in terms of position and momentum operators
i.e $Q=\frac{1}{\sqrt{2}}(a^{\dagger }+a),$
$P=\frac{i}{\sqrt{2}}(a^{\dagger}-a)$, use the displacement
operator property
$D(\frac{s+it}{\sqrt{2}})QD(\frac{s+it}{\sqrt{2}})^{\dagger}=Q-s$
,
$D(\frac{s+it}{\sqrt{2}})PD(\frac{s+it}{\sqrt{2}})^{\dagger}=P-t,$
and the series expansion of the error function, we obtain that
\begin{equation}
D(\frac{s+it}{\sqrt{2}})K(x_{0},k_{0};A,B)D(\frac{s+it}{\sqrt{2}})^{\dagger
}=K(x_{0}+s,k_{0}+t;A,B).
\end{equation}

  Interpreting these two formulas we say that the
family of region operators resulting from rigidly shifting the
vertices of the initial rectangle support i.e $\ S(x_{0},k_{0}
;A,B)\longrightarrow S(x_{0}+s,k_{0}+t;A,B)$, associates itself
with region operators that are unitarily equivalent to the region
operator of the original rectangle $K(x_{0},k_{0};A,B).$ This
implies that all these operators have the same spectrum and in
turn means that their respective eigenvalues i.e the
quasiprobability volumes are equal \rule{5pt}{5pt}

\section{Operational Construction of QPMs: The Case of Canonical Polygons}

Let us introduce rotation operators in terms of the number
operator $N$, to be the operators $R(\frac{2\pi
j}{M})=\exp(\frac{2\pi i}{M}jN),$ which generate rotations by
angles $\frac{2\pi j}{M},$ $j=0,1,...,M-1.$ Let us introduce an
$M-$sided canonical polygon with radial parameter $a,$ defined as
the distance between the center of the polygon and the midpoint of
a side. Due to its geometric invariance under rotation by angles
$\frac{2\pi j}{M},$ $j=0,1,...,M-1,$ the associated polygon region
operator denoted by $X_{[a,M]}$ is obtained as the sum of copies
of the region operator $X_{[a,\frac{2\pi}{M}]}$ of an isosceles
triangle, rotated by all angles $\frac{2\pi j}{M},$
$j=0,1,...,M-1,$ \cite{james}, i.e
\begin{equation}
\varepsilon(X_{[a,\frac{2\pi}{M}]})\equiv X_{[a,M]}=%
{\displaystyle\sum\limits_{j=0}^{M-1}}
R(\frac{2\pi j}{M})X_{[a,\frac{2\pi}{M}]} R(\frac{2\pi j}%
{M})^{\dagger}.
\end{equation}

In the above equation the completely positive trace increasing map
(CPTI) $\varepsilon$, has been introduced by means of its $M$
generators, given by the rotation operators. Since the trace of a
region operator equals the area of its associated region, the CPTI
map $\varepsilon$ that produces an $M-$polygon operator from its
generating isosceles triangle region operator, should be a trace
(area) increasing one. Indeed we have
$Tr\varepsilon(X_{[a,\frac{2\pi
}{M}]})=TrX_{[a,M]}=MTrX_{[a,\frac{2\pi}{M}]}.$ $\ $ More
explicitly we introduce the linear real function of domain
$Dn:RO\rightarrow\mathbf{R},$ defined on any region operator
$X_{A}\in RO,$ determined by a domain $A\in\Gamma,$ of the phase
space, as the area of that domain. Then we obtain from the last
equation the total area of the $M-$gon as $Dn[\varepsilon
(X_{[a,\frac{2\pi}{M}]})]=\bigcup\limits_{j=0}^{M-1}R(\frac{2\pi
j}{M}).[a,\frac{2\pi}{M}]=[a,M]$, where $[a,\frac{2\pi}{M}]$ and
$[a,M]$ parametrize  respectively the triangle and the polygon,
and the notation $R(\frac{2\pi j}{M}).[a,\frac{2\pi}{M}],$
signifies the area swept out from the clockwise rotation of the
triangle $[a,\frac{2\pi}{M}],$ by the angle $\frac{2\pi j}{M}.$

Let us consider also the dual $\varepsilon^{\ast}$ of the CPTI map
$\varepsilon,$ that operates on the density matrix and produces
the same qpi. We recall that for some $\rho$ density state
operator the integral of its associated Wigner function
$W_{\rho}(x,p)$ over the area $S_{a,M}$ of a $a-$canonical $M-$gon
gives the qpi
$Q_{a,M}[W_{\rho}]=\int_{S_{a,M}}W_{\rho}(x,p)\frac{dxdp}{\pi}$.
The dual map $\varepsilon^{\ast}$ is defined as follows
\begin{eqnarray}
Q_{a,M}[W_{\rho}]=Tr(\rho
X_{a,M})\equiv\left\langle\rho,X_{a,M}\right\rangle  =\left\langle
\rho,\varepsilon(X_{a,\frac{2\pi}{M}})\right\rangle =\left\langle
\varepsilon^{\ast}(\rho),X_{a,\frac{2\pi}{M}}\right\rangle
\end{eqnarray}
The bracket in last equation above stems from the observable-state
duality and refers to trace class operators, as the density and
the region operators are considered to be. This last equation
determines the dual CPTI map  $\varepsilon^{\ast}$, which
explicitly reads
\begin{equation}
\varepsilon^{\ast}(\rho)= {\displaystyle\sum\limits_{j=0}^{M-1}}
R(\frac{2\pi j}{M})^{\dagger}\rho R(\frac{2\pi j}{M}).
\end{equation}
Next we seek to construct operationally the polygon region
operator $X_{[a,M]},$ if the triangle region operator
$X_{[a,\frac{2\pi}{M}]}$ is given. Since the maps
$\varepsilon^{\ast},\varepsilon,$ \ are not easy to be construct
physically we can use the standard Naimark extension theorem
\cite{kr}, which applied to CP trace preserving maps allows to
implement them by means of some unitary operator in a space that
is an extension of the original one by an auxiliary (ancilla)
space, followed by a projection to the initial space by means of a
partial trace. This standard unitary dilation theorem of a CPTP
map, will be adapted here to our case for the CPTI dual maps. To
this end we introduce an $M$ dimensional ancilla space $\
H_{A}=\{\left\vert m\right\rangle ,m=0,1,...,M-1\}$ and choose
some operator $\left\vert k\right\rangle \left\langle k\right\vert
$ of it. Then on the total space $H\otimes H_{A},$ we define an
operator $V,$ by means of which we write for the maps
$\varepsilon,$ $\varepsilon^{\ast}$ the respective expressions

\begin{eqnarray}
\varepsilon(X_{[a,\frac{2\pi}{M}]})=X_{[a,M]}=Tr_{A}V(\left\vert
k\right\rangle \left\langle k\right\vert \otimes X_{[a,\frac{2\pi}{M}%
]})V^{\dagger},\qquad \varepsilon^{\ast}(\rho)
=Tr_{A}V^{\dagger}(\left\vert k\right\rangle \left\langle
k\right\vert \otimes\rho)V,
\end{eqnarray}
where we choose $\left\langle j\right\vert V\left\vert
k\right\rangle =R(\frac{2\pi j}{M}),$ $j=0,1,\ldots,M-1.$

In physical terms, this realization describes the coupling of the
given quantum density matrix $\rho$ or the quantum observable
$X_{[a,\frac{2\pi }{M}]}$, with an ancilla, "atomic"
$M-$dimensional system, their dynamical interaction by means of
the operator $V,$, and their subsequent decoupling as a result of
partially tracing the ancilla system. The physical resources of
such an operational realization is however very demanding since
they require the interaction with a high dimensional "atomic"
system. More feasible realizations which are also motivated
geometrically by the various ways of covering a polygon by rotated
and reflected triangles as desired. In fact one would like to have
realization that use only two dimensional ancilla systems i.e
qubits. Although there is no explicit proof detail search shows
that there are no realizations that would use only qubits.

In fact alternative realizations would involve multiple CP maps
and Kraus generators\cite{kr}. Those generators would be the
identity map $\mathbf{1}$, a rotation operator $R(\varphi)$ of
angle $\varphi,$ or a reflection operator $A(r)$ with respect to
some axis e.g going through the origin and making an angle $r$,
with the x-axis. Two types of maps operating on quantum
observables are then introduced, one for rotations and one for
reflections, which read respectively
\begin{eqnarray}
\varepsilon_{\varphi}(X)=X+R(\varphi)XR(\varphi)^{\dagger},\qquad
\varepsilon_{r}(X)=X+A(r)XA(r)^{\dagger}.
\end{eqnarray}
For observables that are region operators corresponding to a phase
space domain with parameters specified by $D$, we denote the
region operator by $X_{D}.$ Then the linearity of CP map yields
for the transformed, by rotation and reflection
maps, region operators the respective relations $\varepsilon_{\varphi}%
(X_{D})=X_{D\cup R(\varphi).D},$ and
$\varepsilon_{r}(X_{D})=X_{D\cup A(r).D}.$ The notation $D\cup
R(\varphi).D$ ($D\cup A(r).D),$ denotes that the domain
Dn[$\varepsilon_{\varphi}(X_{D})$]$\equiv D\cup R(\varphi).D,$
(Dn[$\varepsilon_{r}(X_{D})$]$\equiv D\cup A(r).D),$ of
transformed, by the rotation (reflection) CP map, region operator,
is the union of the domain of the untransformed one by the rotated
$R(\varphi).D$ (reflected $A(r).D$) image of it. This geometric
property common to all CPTI maps, would probably justify the name
"copy-paste transformation" for them. Dually we consider next the
CP map $\varepsilon_{k}^{\ast}$ , $k=\varphi,r,$ with respective
Kraus generators $S=R(\varphi),$ $A(r),$ acting on some
density operator. The introduction of an ancilla Hilbert space $H_{A}%
=span\{\left\vert \Phi\right\rangle ,\left\vert
\Phi^{\bot}\right\rangle \},$ with orthogonal basis vectors i.e
$\langle\Phi|\Phi^{\perp} \rangle=0$, where $\left\vert
\Phi\right\rangle =a\left\vert 0\right\rangle +b\left\vert
1\right\rangle ,$ allows one to implement those CP maps unitarily as
\begin{equation}
\varepsilon_{k}^{\ast}(\rho)=\rho+S^{\dagger}\rho
S=Tr_{A}(V_{k}\left\vert \Phi\right\rangle \left\langle
\Phi\right\vert \otimes\rho V_{k}^{\dagger }).
\end{equation}
The Kraus generators are obtained by the relations ${\mathbf
1}=\left\langle \Phi\right\vert V_{k}\left\vert
\Phi\right\rangle$, and  $S^{\dagger }=\left\langle
\Phi^{\bot}\right\vert V_{k}\left\vert \Phi \right\rangle$. The
unitary operator $V_{k},$ determines given CP map up to a unitary
local (operating non trivially only in ancilla space ) operator
i.e for $W$ a unitary operator in $H_{A},$ the transformation
$V_{k} \longrightarrow W\otimes{\mathbf 1}$ $V_{k},$ generates the
same CP map. In our case such a $V_{k}$ operator in the
$\{\left\vert \Phi\right\rangle ,\left\vert
\Phi^{\bot}\right\rangle \}$ basis is
\begin{eqnarray}
V_{k} & =\left\vert \Phi\right\rangle \left\langle \Phi\right\vert
\otimes{\mathbf 1}+\left\vert \Phi\right\rangle \left\langle
\Phi^{\bot }\right\vert \otimes(-S)+\left\vert
\Phi^{\bot}\right\rangle \left\langle \Phi\right\vert \otimes
S^{\dagger}+\left\vert \Phi^{\bot}\right\rangle
\left\langle \Phi^{\bot}\right\vert \otimes{\mathbf 1}\nonumber\\
& =\sum_{i,j=1,2}\left\vert \Phi^{i}\right\rangle \left\langle \Phi
^{j}\right\vert \otimes V_{k}^{ij}=\left(
  \begin{array}{cc}
    {\mathbf 1}& -S \\
    S^{\dagger} & {\mathbf 1}, \\
  \end{array}
\right)
\end{eqnarray}
where $\{\left\vert \Phi^{1}\right\rangle ,\left\vert \Phi
^{2}\right\rangle \}=$\bigskip$\{\left\vert \Phi\right\rangle
,\left\vert \Phi^{\bot}\right\rangle \},$ and $\
V_{k}^{11}=V_{k}^{22}=\mathbf{1},$ $V_{k}^{21}=-V_{k}^{12}=S$.

Next introducing and combining the next two maps
$\varepsilon_{1}$, $\varepsilon_{1}$, we can construct as an
example, the hexagon region operator with radial parameter
$a=\frac{\sqrt {3}}{2}.$ This realization reads
\begin{equation}
\varepsilon_{2}(\varepsilon_{1}(X_{[a,\frac{2\pi}{6}]}))=X_{[a,6]},
\end{equation}
where, using the equality
$R(\frac{5\pi}{3})=R(\frac{\pi}{3})^{\dagger }=R(-\frac{\pi}{3})$
(see fig.2 also), we have the explicit maps
\begin{eqnarray}
\varepsilon_{1}(X_{[a,\frac{2\pi}{6}]})  =X_{[a,\frac{2\pi}{6}]}%
+R(\frac{\pi}{3})X_{[a,\frac{2\pi}{6}]}R(\frac{\pi}{3})^{\dagger}+R(\frac
{5\pi}{3})X_{[a,\frac{2\pi}{6}]}R(\frac{5\pi}{3})^{\dagger},\\
\varepsilon_{2}(X_{[a,\frac{2\pi}{6}]})  =X_{[a,\frac{2\pi}{6}]}%
+A(\frac{\pi}{2})X_{[a,\frac{2\pi}{6}]}A(\frac{\pi}{2})^{\dagger}.
\end{eqnarray}
The domain of the initial isosceles triangle $[a,\frac{2\pi}{6}]$
increases by successive "copy-paste" induced by the two maps until
reaching the hexagon domain $[a,6]$ as the following equations
show,
\begin{eqnarray}
Dn[\varepsilon_{1}(X_{[a,\frac{2\pi}{6}]})] =[a,\frac{2\pi
}{6}]\cup R(\frac{\pi}{3}).[a,\frac{2\pi}{6}]\cup R(-\frac{\pi}{3}%
).[a,\frac{2\pi}{6}]\nonumber\\
Dn[\varepsilon_{2}(\varepsilon_{1}(X_{[a,\frac{2\pi}{6}]}))]
=Dn[\varepsilon_{1}(X_{[a,\frac{2\pi}{6}]})]\cup A(\frac{\pi}%
{2}).Dn[\varepsilon_{1}(X_{[a,\frac{2\pi}{6}]})]=Dn[X_{[a,6]}]=[a,6].
\end{eqnarray}


See fig.2 for the geometric effect of the combined CPTI maps to
construct the hexagon region operator. An implementation of the
maps based on unitary operators is then given by the relation
\begin{eqnarray}
\varepsilon_{2}(\varepsilon_{1}(X_{[a,\frac{2\pi}{6}]}))  &
=Tr_{A}\otimes Tr_{A}\left(
V_{r=\frac{\pi}{2}\varphi=\pm\frac{\pi}{3}}\left\vert \Theta
^{1}\right\rangle \left\langle \Theta^{1}\right\vert
\otimes\left\vert
\Phi^{1}\right\rangle \left\langle \Phi^{1}\right\vert \right. \nonumber\\
&  \left.  X_{[a,\frac{2\pi}{6}]}V_{r=\frac{\pi}{2}\varphi=\pm\frac{\pi}{3}%
}^{\dagger}\right) \nonumber\\
&  \equiv\sum_{i,j=1,2}\sum_{k,l=1,2}V_{r=\frac{\pi}{2}}^{ij}V_{\varphi
=\pm\frac{\pi}{3}}^{kl}X_{[a,\frac{2\pi}{6}]}V_{\varphi=\pm\frac{\pi}{3}%
}^{\dagger kl}V_{r=\frac{\pi}{2}}^{\dagger ij}%
\end{eqnarray}
where two ancilla Hilbert spaces, one three dimensional i.e $\ $ $H_{A}%
^{1}=span\{\left\vert \Theta^{1}\right\rangle ,\left\vert \Theta
^{2}\right\rangle ,\left\vert \Theta^{3}\right\rangle \},$ and one
two dimensional $H_{A}^{2}=span\{\left\vert \Phi^{1}\right\rangle
,\left\vert \Phi^{2}\right\rangle \}$ have been used. The
operators involved are also of two kinds: the first one
$V_{\varphi=\pm\frac{\pi}{3}}$ is a three dimensional unitary
dilation of the Kraus generators of \ the CPTI "rotation" map
$\varepsilon_{1}$ i.e $\ \mathbf{1=}\left\langle 0\right\vert
V_{\varphi=\pm\frac{\pi}{3}}\left\vert \Theta^{1}\right\rangle ,$
$R(\frac {\pi}{3})=\left\langle 1\right\vert
V_{\varphi=\pm\frac{\pi}{3}}\left\vert \Theta^{1}\right\rangle
,R(\frac{5\pi}{3})=$ $\left\langle 2\right\vert
V_{\varphi=\pm\frac{\pi}{3}}\left\vert \Theta^{1}\right\rangle ,$
with the property
$V_{\varphi=\pm\frac{\pi}{3}}V_{\varphi=\pm\frac{\pi}{3}}^{\dagger
}=3\mathbf{1,}$. The second one $V_{r=\frac{\pi}{2}}$, is also a
two dimensional dilation \ of the "reflection" \ CPTI map
$\varepsilon_{2}$ defined in the canonical basis as
$V_{r=\frac{\pi}{2}}=\left(
\begin{array}
[c]{cc}%
\mathbf{1} & -A(\frac{\pi}{2})^{\dagger}\\
A(\frac{\pi}{2}) & \mathbf{1}%
\end{array}
\right)  ,$ and is such that
$V_{r=\frac{\pi}{2}}V_{r=\frac{\pi}{2}}^{\dagger }=2\mathbf{1.}$
It should be noticed that the reflection with respect to the
y-axis i.e $(x,y)\rightarrow(-x,y),$ is not a canonical
transformation, but since the domain
Dn[$\varepsilon_{1}(X_{[a,\frac{2\pi}{6}]})],$ is the semicircle
we actually need the reflection with respect to the origin i.e
$(x,y)\rightarrow (-x,-y),$ which in fact is a canonical
transformation, and it is realized by the parity operator
$\Pi=e^{i\pi N}=(-1)^{N},$ so we will in effect use the operator
$V_{r=\frac{\pi}{2}}=\left(
\begin{array}
[c]{cc}%
\mathbf{1} & -\Pi\\
\Pi & \mathbf{1}%
\end{array}
\right)  .$ Indeed the parity operator with properties
$\Pi^{\dagger}=\Pi,$ and $\Pi^{2}=\mathbf{1,}$ operates on the
canonical generators as $\Pi a\Pi=-a,$ $\Pi
a^{\dagger}\Pi=-a^{\dagger},$ and on the displacement operator as
$\Pi D_{\alpha}\Pi=D_{-\alpha}.$

\section{Bounds, Tiling and Majorization for QPMs}

Let us start by pointing out the covariance property of the
displaced parity kernel operator $\Delta(q,p)=D(q,p)\Pi
D^{\dagger}(q,p)$ of the region operator under the displacement
operator, which reads
\begin{equation}
D(q^{^{\prime}},p^{^{\prime}})\Delta(q,p)D^{\dagger}(q^{^{\prime}}%
,p^{^{\prime}})=\Delta(q+q^{^{\prime}},p+p^{^{\prime}}).
\end{equation}
We now introduce a CPTI displacement map by the equation
\begin{equation}
\varepsilon_{_{(q\,^{^{\prime}},p^{^{\prime}})}}(X)=X+D(q^{^{\prime}%
},p^{^{\prime}})XD^{\dagger}(q^{^{\prime}},p^{^{\prime}}).
\end{equation}
Two particular such maps are the $q^{^{\prime}}-$west and the $p^{^{\prime}}%
-$north displacement maps defined correspondingly as
\begin{equation}
\varepsilon_{q^{^{\prime}}}^{W}\equiv\varepsilon_{_{(q\,^{^{\prime}},0)}%
},\qquad \varepsilon_{p^{^{\prime}}}^{N}\equiv\varepsilon_{_{(0,p^{^{\prime}%
})}}.
\end{equation}
For a region operator $X_{[q,p]}$ or $X_{[c,c;d]}$ with support
$[q,p]$ a rectangle with sides $q,p$ along $x,y$ axes, or
$[c,c;d]$ a disk of diameter $d$ and center \thinspace at the
point $(c,c),$ the previously introduced domain operator is
respectively
\begin{equation}
Dn[X_{[q,p]}]=[q,p],\qquad Dn[X_{[c;d]}]=[c;d].
\end{equation}
Also we will use the following symbols to denote
$q^{^{\prime}}-$west and the $p^{^{\prime}}-$north displacements,
of respective domains of phase space
\begin{eqnarray}
W_{q^{^{\prime}}}.[q,p]  &  =[q^{^{\prime}}+q,p],\qquad N_{p^{^{\prime}}%
}.[q,p]=[q,p^{^{\prime}}+p],\\
W_{d}.[c,c;d]  &  =[c+d,c;d],\qquad N_{d}.[c,c;d]=[c,c+d;d]
\end{eqnarray}

The CPTI maps that can induce the respective displacement in the
domains of rectangle region operators are
\begin{eqnarray}
\varepsilon_{p^{^{\prime}}}^{N}(X_{[q,p]})  =X_{[q,p]}+D(0,p^{^{\prime}%
})X_{[q,p]}D^{\dagger}(0,p^{^{\prime}}),\\
\varepsilon_{q^{^{\prime}}}^{W}(X_{[q,p]})  =X_{[q,p]}+D(q^{^{\prime}%
},0)X_{[q,p]}D^{\dagger}(q^{^{\prime}},0),
\end{eqnarray}
and similarly for the respective displacements of a $[c,c;d]$ disk,
\begin{eqnarray}
\varepsilon_{p^{^{\prime}}}^{N}(X_{[c,c;d]})
=X_{[c,c;d]}+D(0,p^{^{\prime
}})X_{[c,c;d]}D^{\dagger}(0,p^{^{\prime}}),\\
\varepsilon_{q^{^{\prime}}}^{W}(X_{[c,c;d]})   =X_{[c,c;d]}+D(q^{^{\prime}%
},0)X_{[c,c;d]}D^{\dagger}(q^{^{\prime}},0),
\end{eqnarray}
Indeed we have for the rectangle that
\begin{eqnarray}
Dn[\varepsilon_{p^{^{\prime}}}^{N}(X_{[q,p]})]    =[q,p]\cup N_{p^{^{\prime}%
}}.[q,p]=[q,p]\cup\lbrack q,p^{^{\prime}}+p],\\
Dn[\varepsilon_{q^{^{\prime}}}^{W}(X_{[q,p]})]    =[q,p]\cup W_{q^{^{\prime}%
}}.[q,p]=[q,p]\cup\lbrack q^{^{\prime}}+q,p].
\end{eqnarray}
and for the disk
\begin{eqnarray}
Dn[\varepsilon_{p^{^{\prime}}}^{N}(X_{[c,c;d]})]    =[c,c;d]\cup
N_{p^{^{\prime}}}.[c,c;d]=[c,c;d]\cup\lbrack c,c+p^{^{\prime}};d],\\
Dn[\varepsilon_{q^{^{\prime}}}^{W}(X_{[c,c;d]})]   =[c,c;d]\cup
W_{q^{^{\prime}}}.[c,c;d]=[c,c;d]\cup\lbrack c+q^{^{\prime}},c;d].
\end{eqnarray}

In view of fig.3, we now introduce a rectangle operator
$X_{[\alpha,\beta]},$ (fig.3a), by $\varepsilon_{\beta}^{W}$ along
the $x$ axis to west, (fig.3b), and one displacement of the
resulting rectangle by $\varepsilon_{\alpha}^{N}$ along the $y$
axis to north, (fig.3c). The combined displacements define the map
$\varepsilon_{(\alpha,\beta)}$ realizing a single step in the
tiling by operating on the square and disk region operators
respectively as follows
\begin{eqnarray}
X_{[2\alpha,2\beta]}
=\varepsilon_{(\alpha,\beta)}(X_{[\alpha,\beta
]})\equiv\varepsilon_{\beta}^{N}\circ\varepsilon_{\alpha}^{W}(X_{[\alpha
,\beta]}),\\
X_{[c+d,c+d;d]}  =\varepsilon_{(d,d)}(X_{[c,c;d]})\equiv\varepsilon_{d}%
^{N}\circ\varepsilon_{d}^{W}(X_{[c,c;d]}),
\end{eqnarray}
where the notation $X_{[c+d;c+d;d]}$ means the cluster of disks of
diameter $d,$ with center at the points
$\{(c,c);(c,c+d);(c+d,d);(c+d,c+d)\}.$ This notation extended in the
form $X_{[c+md,c+md;d]},$ where $m=0,1,2,...$ , describes the
cluster of all disks of diameter $d,$ "under" the disk with center
at the point ($c+md,c+md).$

Referring to fig.3 (fig.4) we introduce a west-north tiling of an
initial rectangle (disk) region operator $X_{[\alpha,\beta]},$
$(X_{[c,c;d]}),$ by successively operating on it with the CPTI
maps $\{\varepsilon_{(\alpha
,\beta)},\varepsilon_{(2\alpha,2\beta)},\varepsilon_{(4\alpha,4\beta
)},...\},(\{\varepsilon_{(d,d)},\varepsilon_{(2d,2d)},\varepsilon
_{(4d,4d)},...\}),$ which results into the sequence of region
operators
$\{X_{[\alpha,\beta]},X_{[2\alpha,2\beta]},X_{[3\alpha,3\beta]},...\},$
$(\{X_{[c,c;d]},X_{[c+d,c+d;d]},X_{[c+2d,c+2d;d]},...\}),$ the
sequence of domains of which corresponds to a rectangle (disk)
shaped west-north tiling
of the upper left part of the plane. In other words we have the iteration%
\begin{eqnarray}
X_{[2\mu,2\nu]}    =\varepsilon_{(\mu,\nu)}(X_{[\mu,\nu]}),\qquad%
\mu=m\alpha,\qquad \nu=m\beta,\qquad m\in\{2,4,6,...\},\\
X_{[c+md,c+md;d]}    =\varepsilon_{((m-1)d,(m-1)d)}(X_{[c+(m-1)d,c+(m-1)d;d]}%
),\qquad m\in\{1,2,3,...\},
\end{eqnarray}
for the two types of tiling of the original rectangle and disk
region operators.

Before proceeding let's us mention that the CPTI maps can be
unitarily extended by utilizing an auxiliary two dimensional Hilbert
space and a partial tracing as follows ($y=$rectangle, disk),
\begin{eqnarray}
\varepsilon_{q^{^{\prime}}}^{W}(X_{[y]})    =Tr_{A}(U_{q^{^{\prime}}}%
^{W}|0\rangle\langle0|\otimes
X_{[y]}U_{q^{^{\prime}}}^{W\dagger}),\qquad \varepsilon_{p^{^{\prime}}}^{N}(X_{[y]})  =Tr_{A}(U_{p^{^{\prime}}}%
^{N}|0\rangle\langle0|\otimes X_{[y]}U_{p^{^{\prime}}}^{N\dagger}).
\end{eqnarray}
The unormalized unitary operators chosen  for the extension of the
west and north displacement maps are respectively the following
two
\begin{equation}
U_{q^{^{\prime}}}^{W}=\left(
\begin{array}
[c]{cc}%
\mathbf{1} & -D^{\dagger}(q^{^{\prime}},0)\\
D(q^{^{\prime}},0) & \mathbf{1}%
\end{array}
\right)  ,\qquad U_{p^{^{\prime}}}^{N}=\left(
\begin{array}
[c]{cc}%
\mathbf{1} & -D^{\dagger}(0,p^{^{\prime}})\\
D(0,p^{^{\prime}}) & \mathbf{1}%
\end{array}
\right)  .
\end{equation}

Let us turn now to the question of determining the quasi
probability masses (qpm) identified with the eigenvalues of
rectangle (disk) region operator
$X_{[2\mu,2\nu]}\\(X_{[c+md,c+md;d]})$ at some time $[2\mu,2\nu]$
$(m)$, of the tiling iteration process, from its previous one
$[\mu,\nu]$ $(m-1).$ This would be an important relation for
determining qpm's, and their bounds i.e the maximal and minimal
ones among them, in the course of rectangle (disk) tiling, once
the qpm's of the initial rectangle (disk) region operator is given
or can be measured or estimated. To this end let us take the
canonical decomposition of the matrices of a region operator
$X=V^{\dagger}X^{d}V,$ before and after a general tiling CPTI map
i.e of the
operator $X^{^{\prime}}=\varepsilon_{_{(q\,^{^{\prime}},p^{^{\prime}})}%
}(X)=W^{\dagger}X^{^{d  \prime}}W.$ Here $V$, $W$ and $X^{d},$
$X^{^{d  \prime}},$ are the respective diagonalizing unitary
matrices and the diagonal matrices of the eigenvalues appearing in
the canonical decomposition. The latter two are related by the
equation
\begin{equation}
X^{d\prime}=W[V^{\dagger}X^{d}V+D(q^{^{\prime}},p^{^{\prime}%
})V^{\dagger}X^{d}VD^{\dagger}(q^{^{\prime}},p^{^{\prime}})]W^{\dagger}%
\end{equation}
Let $\lambda^{^{\prime}}=(\lambda_{j}^{^{\prime}%
})_{j=1}^{\infty}=((X^{^{d \prime}})_{jj})_{j=1}^{\infty},$ the
eigenvalue column vector, and \ similarly from the eigenvalues of
the $X^{d}$ the vector $\lambda.$ In view of the last equation these
two real column vectors are connected by the following matrix
$\Sigma$
\begin{equation}
\lambda^{^{\prime}}=\left(  WV^{\dagger}\circ\overline{WV^{\dagger}%
}+WD(q^{^{\prime}},p^{^{\prime}})V^{\dagger}\circ\overline{WD(q^{^{\prime}%
},p^{^{\prime}})V^{\dagger}}\right)  \lambda\equiv\Sigma\lambda.
\end{equation}
Above, the overbar denotes complex conjugation and, the
elementwise or Hadamard product between matrices defined as
$(A\circ B)_{ij}=A_{ij}B_{ij},$ has been used. Since for a given
unitary matrix $U$ the Hadamard product $U\circ \overline{U}$ is a
column and row stochastic matrix (bistochastic), i.e for the unit
column vector $e=(1,1,...)^{T},$ we have that $U\circ\overline
{U}e=e,$ and $e^{T}U\circ\overline{U}=e^{T},$ then our present
sigma matrix has column and row sums equal to two i.e $\Sigma
e=2e,$ and $e^{T}\Sigma=2e^{T}.$

In order to apply these facts to two successive region operators
in the rectangle tiling, we introduce first their canonical
decompositions
\begin{eqnarray}
\bigskip X_{[2\mu,2\nu]}   =W^{\dagger}X_{[2\mu,2\nu]}^{d}W=W^{\dagger
}diag(\lambda_{\lbrack2\mu,2\nu]})W,\\
X_{[\mu,\nu]} =V^{\dagger}X_{[\mu,\nu]}^{d}V=V^{\dagger}diag(\lambda
_{\lbrack\mu,\nu]})V.
\end{eqnarray}
Then we obtain the following relation among their diagonal
components
\begin{eqnarray}
X_{[2\mu,2\nu]}^{d}
=W\{V^{\dagger}X_{[\mu,\nu]}^{d}V+D(\mu,0)V^{\dagger
}X_{[\mu,\nu]}^{d}VD(\mu,0)^{\dagger}\nonumber\\
+D(0,\nu)V^{\dagger}X_{[\mu,\nu]}^{d}VD(0,\nu)^{\dagger}\nonumber\\
+D(0,\nu)D(\mu,0)V^{\dagger}X_{[\mu,\nu]}^{d}VD(\mu,0)^{\dagger}%
D(0,\nu)^{\dagger}\}W^{\dagger}.
\end{eqnarray}
The diagonal elements of these matrices are the column vectors
$\lambda_{\lbrack\mu,\nu]},$ $\lambda_{\lbrack2\mu,2\nu]},$ and
they are identified with qpm's before and after the tiling map.
They are explicitly related as follows
\begin{eqnarray}
\lambda_{\lbrack2\mu,2\nu]}   =\left(  WV^{\dagger}\circ\overline
{WV^{\dagger}}+WD(\mu,0)V^{\dagger}\circ\overline{WD(\mu,0)V^{\dagger}}\right.
\nonumber\\
 \left. +WD(0,\nu)V^{\dagger}\circ\overline{WD(0,\nu)V^{\dagger}}%
+WD(\mu,\nu)V^{\dagger}\circ\overline{WD(\mu,\nu)V^{\dagger}}\right)
\lambda_{\lbrack\mu,\nu]}\nonumber\\
\equiv\Gamma\lambda_{\lbrack\mu,\nu]}.
\end{eqnarray}
The matrix gamma introduced above updates the vector of qpm's in
the iteration of rectangle tiling, and it has column and row sum
equal to four i.e $\Gamma e=4e,$ $e^{T}\Gamma=4e^{T},$. This last
fact signifies the quadrupling of a initial rectangle at each
tiling step. As to the disk tiling, the analogous relation results
after simplifying by setting $W=\mathbf{1,}$ since the disk range
operator $X_{[c,c;d]}$
is diagonal in the number state basis i.e $X_{[c,c;d]}=X_{[c,c;d]}^{d}%
=diag(\lambda_{\lbrack c,c;d]}).$ Then the relation connecting the column
vectors $\lambda_{\lbrack c,c;d]}$ and $\lambda_{\lbrack c+d,c+d;d]},$ of the
diagonal elements of the respective disk range operators $X_{[c+d,c+d;d]}%
=\varepsilon_{(d,d)}(X_{[c,c;d]})=V^{\dagger}diag(\lambda_{\lbrack
c+d,c+d;d]})V,$ and $X_{[c,c;d]},$ i.e the operators before and
after a disk tiling map, reads
\begin{eqnarray}
\lambda_{\lbrack c+d,c+d;d]}   =\left(  V^{\dagger}\circ\overline
{V^{\dagger}}+D(d,0)V^{\dagger}\circ\overline{D(d,0)V^{\dagger}}\right.
\nonumber\\
 \left.  +D(0,d)V^{\dagger}\circ\overline{D(0,d)V^{\dagger}}
+D(d,d)V^{\dagger}\circ\overline{D(d,d)V^{\dagger}}\right)  \lambda_{\lbrack
c,c;d]}\equiv E\lambda_{\lbrack c,c;d]}.
\end{eqnarray}
As with the matrix $\Gamma,$ the matrix $E$ has column and row sums
equal to four.

\textit{Remarks:} 1) The eigenvalues $\lambda_{\lbrack2\mu,2\nu]}$
i.e the qpm's, will result if we evaluate the matrix elements of
some square region operators in some of their eigenvectors. These
eigenvectors however may not be known and not even been relevant
vectors for some problems, in such cases the above recurrent
relations may not be useful. Instead the determination of matrix
elements in the number states which are physically relevant
states, is desirable and this will be pursued next.

2) The same remark is in fact true for the determination of
eigenvalues $\lambda _{\lbrack c+md,c+md;d],}$ i.e the qpm's in
the case of disk tiling. So in the following we evaluate the
diagonal matrix elements of the respective region operators in the
number states.

\begin{theorem}{Lemma} \label{TH4}
Let the transformation $D\rightarrow\varepsilon(D)\equiv
X=\int_{x\in\Delta
}V(x)DV^{\dagger}(x)d\mu(x),$ of the operator $D=\sum_{k=0,1,...}%
d_{k}|k\rangle\langle k|,$ diagonal in the number state basis, and
$V=V(x)$ a unitary operator depending on some variable $x$ with
values in some domain $\Delta,$ on which a integration measure
$d\mu(x)$ is defined. Let the vector
$|\Psi\rangle=\sum_{k=0,1,...}d_{k}|k\rangle,$ then the diagonal
elements of operator $X,$ are determined by $\langle
k|X|k\rangle=\langle k|\left(  \int_{x\in\Delta}V(x)\circ
\overline{V(x)}d\mu(x)\right)  |\Psi\rangle$.
\end{theorem}
The proof is straightforward, and the lemma is also true for the
case of transformations defined by discrete sums. All the above can
now be summarized in the following proposition:
\begin{theorem}{Proposition} \label{TH5}
Let the rectangle and disk tiling with iteration maps
$\varepsilon_{(\mu ,\nu)}$ and $\varepsilon_{(md,md)}$
respectively, acting on rectangle and disk region operator as
$\varepsilon_{(\mu,\nu)}(X_{[\mu,\nu]})=X_{[2\mu,2\nu ]}$ and
$\varepsilon_{(md,md)}(X_{[c+md,c+md;d]})=X_{[c+(m+1)d,c+(m+1)d;d]}$.
Let the canonical decomposition of the rectangle tiling range
operator
$X_{[\mu,\nu]}=V^{\dagger}X_{[\mu,\nu]}^{d}V=V^{\dagger}diag(\lambda
_{\lbrack\mu,\nu]})V,$ with
$\lambda_{\lbrack\mu,\nu]}=\{\lambda_{\lbrack
\mu,\nu]}^{k}\}_{k=0}^{\infty},$ the column vector of its
eigenvalues. Let $X_{[c+md,c+md]}
=V^{\dagger}X_{[c+md,c+md]}^{d}V=V^{\dagger}diag(\lambda_{\lbrack
c+md,c+md]})V$ been also the canonical decomposition of the disk
tiling range operator with $\lambda_{\lbrack
c+md,c+md]}=\{\lambda_{\lbrack c+md,c+md]} ^{k}\}_{k=0}^{\infty}$
the column vector of its eigenvalues. Given that the
eigenvalue-vector problem has been solved for these operators
after some given tiling step, then their diagonal elements in the
number state basis, (which physically signify the quasiprobability
mass of the Wigner function in the respective number states), are
given by the two corresponding expressions,
\begin{eqnarray}
\langle k|X_{[2\mu,2\nu]}|k\rangle   =\langle k|\left[  V^{\dagger}%
\circ\overline{V}+D(\mu,0)V^{\dagger}\circ\overline{VD(\mu,0)^{\dagger}%
}+D(0,\nu)V^{\dagger}\circ\overline{VD(0,\nu)^{\dagger}}\right. \nonumber\\
 \left.
+D(\mu,\nu)V^{\dagger}\circ\overline{VD(\mu,\nu)^{\dagger}}\right]
|\Phi_{\lbrack\mu,\nu]}\rangle,
\end{eqnarray} and
\begin{eqnarray}
\langle k|X_{[c+(m+1)d,c+(m+1)d;d]}|k\rangle  =\langle k|\left[
V^{\dagger
}\circ\overline{V}+D(md,0)V^{\dagger}\circ\overline{VD(md,0)^{\dagger}}\right.
\nonumber\\
+D(0,md)V^{\dagger}\circ\overline{VD(0,md)^{\dagger}
}\nonumber\\
\left. +D(md,md)V^{\dagger}\circ\overline{VD(d,md)^{\dagger}}\right]
|\Psi_{\lbrack c+md,c+md]}\rangle.
\end{eqnarray}
where
$|\Phi_{\lbrack\mu,\nu]}\rangle=\sum_{k=0,1,...}\lambda_{\lbrack
\mu,\nu]}^{k}|k\rangle$ and $|\Psi_{\lbrack
c+md,c+md]}\rangle=\sum_{k=0,1,...}\lambda_{\lbrack
c+md,c+md]}^{k}|k\rangle$.
\end{theorem}

The description of the tiling process that has been given above
with rectangle and disc tiles is based respectively on the one
step CPTI maps $\
\varepsilon_{(\mu,\nu)}(X_{[\mu,\nu]})=X_{[2\mu,2\nu]},\ \ $and
$\varepsilon_{(md,md)}(X_{[c+md,c+md;d]})=X_{[c+(m+1)d,c+(m+1)d;d]}.$
These maps induce in the eigenvalue vectors of the region
operators the corresponding transformations $\lambda_{\lbrack
c+d,c+d;d]}=E\lambda_{\lbrack c,c;d]},$ and
$\lambda_{\lbrack2\mu,2\nu]}=\Gamma\lambda_{\lbrack\mu,\nu]}.$ The
matrices $E$ and $\Gamma$ involved, are almost bistochastic, since
as we have shown they have column and raw sums equal to four. This
property allows to estimate the behavior of upper and lower bounds
of the eigenvalues of the region operators after a one step
operation of the tiling process. Specifically in the subsequent
corollary we will determine the relation before and after the
performance of one step tiling, between the extremal values of the
eigenvalues of rectangle and disk region operators. These values
of course stand for the extremal values of the quasiprobability
mass of the respective region operators in their associated
eigenstates. These extremal values provide useful information by
means of which we can estimate the interval of values of the qpm's
in the course of tiling process.

To this end let us introduce the following notation: let the
matrix $S$ which stands for the matrices $E$ and $\Gamma$ used
above, sharing the same properties. This matrix will be used to
write generically $\lambda^{^{\prime}}=S\lambda,$ for a
transformation of the eigenvalues in a one-step tiling. Let us
introduce the \textit{non-increasing} (\textit{non-decreasing})
ordering of the eigenvalue vectors i.e respectively
$\lambda^{\downarrow}=(\lambda_{1}^{\downarrow
},\lambda_{2}^{\downarrow},...,\lambda_{i}^{\downarrow},\lambda_{i+1}%
^{\downarrow},...),$ with $\lambda_{i}^{\downarrow}\geq\lambda_{i+1}%
^{\downarrow},$ $i=1,2,...$ , and
$\lambda^{\uparrow}=(\lambda_{1}^{\uparrow
},\lambda_{2}^{\uparrow},...,\lambda_{i}^{\uparrow},\lambda_{i+1}^{\uparrow
},...),$ with $\lambda_{i}^{\uparrow}\leq\lambda_{i+1}^{\uparrow},$
$i=1,2,...$ .

\begin{theorem}{Corollary}
Let the transformation $\lambda^{^{\prime}}=S\lambda,$ of the
column vector of rectangle/disk region operator eigenvalues in a
one-step tiling map. For the non-increasing and the non-decreasing
eigenvalue vectors the following respective majorization relations
are valid, $4\lambda^{\downarrow }\succ\lambda^{\downarrow\prime}$
and $\lambda^{\uparrow/}\succ4\lambda ^{\uparrow}.$ Also for the
maximum and minimum values of the eigenvectors before and after
the transformation the "\textit{squeezing" property is valid
}$4\lambda_{\min}\leq\lambda_{\min}^{^{\prime}}<\lambda_{\max}^{^{\prime}}%
\leq4\lambda_{\max}.$
\end{theorem}

\textit{Proof:} We start with the equation
$\lambda^{^{\prime}}=S\lambda,$ but it suffices to assume
$\lambda^{^{^{\downarrow}\prime}}=S\lambda ^{^{\downarrow}},$ and
$\lambda^{\uparrow^{\prime}}=S\lambda^{\uparrow},$ for we could
always consider two suitable permutations $P,$ $Q$ such that
$\lambda ^{^{^{\downarrow}\prime}}=P\lambda^{^{\prime}},$
$\lambda^{^{^{\downarrow}} }=P\lambda.$ The equation for the
non-increasing ordered vectors
$\lambda^{^{^{\downarrow}\prime}}=PSW^{T}\lambda^{^{^{\downarrow}}},$
involves the matrix $PSW^{T}$ with the same property as the $S,$
matrix i.e $PSW^{T}e=PSe$ $=4Pe=4e$ and
$e^{T}PSW^{T}=e^{T}SW^{T}=4e^{T}W^{T}=4e^{T}.$ Similar result are
obtained for the non-decreasingly ordered vectors. We start now
with the equation $\lambda^{^{^{\downarrow}\prime}}
=S\lambda^{^{\downarrow}}$ and write
$\lambda^{^{^{\downarrow}\prime}}
=4\frac{S}{4}\lambda^{^{\downarrow}}\equiv4H\lambda^{^{\downarrow}},$
where $H$ is a bistochastic matrix. The transformation
$\widetilde{\lambda }^{^{\downarrow}}\equiv
H\lambda^{^{\downarrow}},$ leads to the majorization
$\lambda^{^{\downarrow}}\succ\widetilde{\lambda}^{^{\downarrow}}$
\cite{mo}. Then the relation $\frac{1}{4}\lambda^{^{^{\downarrow
}\prime}}=\widetilde{\lambda}^{^{\downarrow}},$ leads to the
majorization
$4\lambda^{\downarrow}\succ\lambda^{\downarrow\prime}.$ The proof
of the second majorization relation is similar. Finally, taking
the maximum and minimum values of qpm's before i.e $\lambda_{\max
}=\lambda_{1}^{^{\downarrow}}$,
$\lambda_{\min}=\lambda_{1}^{\uparrow},$ and after i.e
$\lambda_{\max}=\lambda_{1}^{^{\downarrow}}$,
$\lambda_{\min}=\lambda _{1}^{\uparrow},$ of the tiling map, and
using the inequalities ensuing from the majorization relations we
obtain the "squeezing" \ property $4\lambda
_{\min}\leq\lambda_{\min}^{^{\prime}}<\lambda_{\max}^{^{\prime}}\leq
4\lambda_{\max}$\rule{5pt}{5pt}

\section{Conclusions}
Region operators (RO) associated with integrals of Wigner
quasiprobability function over domains of phase space provide an
example of new type of quantum mechanical observables. These
operators can be considered as a quantum version of characteristic
functions of classical regions, or as generalized operator valued
probability measures (OVM). The theory connecting geometric
transformations of regions to spectral characteristics of ROs, has
been initiated in this work. Applications of ROs and their
associated qpm's with their respected bounds, to the field of
tomographic reconstruction of Wigner function over certain regions
\cite{leo,sch}, should be a testing ground for the theory. Also
the application of region operators in the study of coupled
bipartite quantum systems, could provide a framework for
investigation of the geometric manifestation and quantification of
quantum entanglement, cast in the language of generalized operator
valued measures on phase space. We aim to return to these matters
elsewhere.

\pagebreak

\vskip 1cm
 Figure captions.\\ Fig.1 Point and line region operators
$K_0=\Pi$ and
$K_0^P=\varepsilon_L^P(\Pi)$ respectively. \\

Fig.2 Construction of hexagon region operators $X_{[a,6]}$ (2c), by
implementation of two CPTI maps on the
isosceles triangle region operator $X_{[a,\frac{\pi}{3}]}$ (2a).\\

Fig.3 Construction of west-north tiling by successive application of
CPTI maps on the rectangle region operators $X_{[\alpha,\beta]}$ (3a).\\

Fig.4 Construction $(i)$ of west-north (4b)and $(ii)$ of
reflection-rotation (4d, 4e) tiling by successive application of
CPTI maps on region operators of clusters of disks (4a) and
rectangles  (4c) respectively.


\begin{thebibliography}{99}

\bibitem {schleich} W. P. Schleich,
{\em Quantum Optics in Phase Space}, Wiley-VCH, Berlin 2001.
\bibitem {leo}U. Leonhardt, {\em Measuring the Quantum State of
Light}, Cambridge University Press, Cambridge 1997.
\bibitem {sch}W. P. Schleich and M. G. Raymer: {\em J.\ Mod.\ Opt.} {\bf 44} 12
(1997).
\bibitem {bdw} A. J. Bracken, H-D. Doebner and J. G. Woods: {\em Phys.\ Rev.\ Lett.} {\bf 83}, 3758
(1999);(arXiv.org, quant-ph/9905097).
\bibitem {bew1}
A. J. Bracken, D. Ellinas and J.G. Wood, {\em Acta \ Phys. \
Hungarica B, \ Quantum \ Electronics} {\bf 20} 121
(2004);(arXiv.org, quant-ph/0304110).
\bibitem {bew2} A. J. Bracken, D. Ellinas and J. G. Woods {\em J.\ Phys.\ A} {\bf 36}, L297
(2003);(arXiv.org, quant-ph/0304010).
\bibitem {james}J. G. Woods, PhD Thesis, University of Queensland,
Brisbane, 2004.
\bibitem {kr}K. Kraus, {\em States, Effects and Operations}, Springer, Berlin 1983.
\bibitem {woot}W. K. Wootters, {\em Ann.\ Phys.} {\bf 176}, 1
(1987).
\bibitem {mo}A. W. Marshall and  I. Olkin {\em Inequalities: Theory of Majorization and Its
Applications}, Academic Press, New York 1979.
\bibitem {iwop}Fan Hong-Yi, H. R. Zaidi and J. R. Klauder: {\em Phys.\ Rev.\
D} {\bf 35}, 1831--1831 (1987), and subsequent works by the first
author.
\end{thebibliography}
\end{document}